![DTIP 2007 of MEMS & MOEMS] Stresa, Italy, 25-27 April 2007# A HIGH POWER DENSITY ELECTROSTATIC VIBRATION-TO-ELECTRIC ENERGY CONVERTER BASED ON AN IN-PLANE OVERLAP PLATE (IPOP) MECHANISM

*Ayyaz Mahmood Paracha, Philippe Basset, Frédéric Marty,
Adrian Vaisman Chasin, Patrick Poulichet, Tarik Bourouina*

ESIEE-ESYCOM, BP-99, 2 bd Blaise Pascal, 93162 Noisy-le-Grand Cedex, France

(Corresponding author: Philippe Basset, email: p.basset@esiee.fr, tel: +33 1 45 92 65 86)**ABSTRACT**

In this paper, design, fabrication and characterization issues of a bulk silicon-based, vibration powered, electric energy generator are addressed. The converter is based on an In-Plane Overlap Plate (IPOP) configuration [1]. Measurements have shown that with a theoretically lossless electronics and a starting voltage of 5 V, power density of 58 µW/cm$^3$ is achievable at the resonance frequency of 290 Hz. It can be further improved by reducing the parasitic capacitance, which can be achieved by silicon etching, but a considerable mass is lost. In [2], it is shown that 19% of mass reduction improves power density from 12.95 µW/cm$^3$ to 59 µW/cm$^3$. Hence an enhancement in fabrication process is proposed, which is termed as Backside DRIE. It helps in increasing power density without loosing an important quantity of mass. Simulations have shown that 2.5% of mass removal improves power density up to 76.71 µW/cm$^3$. Initial simulation results and problems of associated electronics are also discussed.## 1. INTRODUCTION

Various efforts have already been made to reduce the power consumption of sensors, and to make them autonomous i.e. independent of externally attached power sources [3]. Hence energy should be harvested from the ambient environment to power these sensors. This concept has lot of advantages: number of wires used for the electrical connections in between the main power source and different sensors can be reduced, device's lifetime increases, cost maintenance decreases as there is no need to replace the external batteries. In addition, various chemical materials, currently being used in power sources, can be eliminated. In order to provide autonomous nodes in a sensor network (cf. Fig. 1) and to power smart-dusts [4], ambient vibration is a power source that is largely considered [5], as there exists broadband vibration spectrum in applications like automotives [6], air planes, etc…

The three main mechanisms to harvest energy from environmental ambient vibrations are electrostatic [1, 3, 6], piezoelectric [7] and electromagnetic [8]. There are no definite rules to make a choice among these mechanisms, as it all depends on an application's requirement. But still CMOS compatible MEMS converters, with a high power generation, are a challenge. Among above three defined types, electrostatic-based energy harvesters are more compatible with CMOS process and are considered good for miniaturization.

Electrostatic converters need two sets of electrodes. First one is attached to a moving mass called as proof mass and second one is fixed to the substrate. A huge proof mass is normally used to target a low resonance frequency. External vibrations force the mobile part to move relatively to the substrate, which results in a mechanical-to-electrical energy transduction, if a constant charge or voltage is maintained on the electrodes while the capacitance decreases [9].

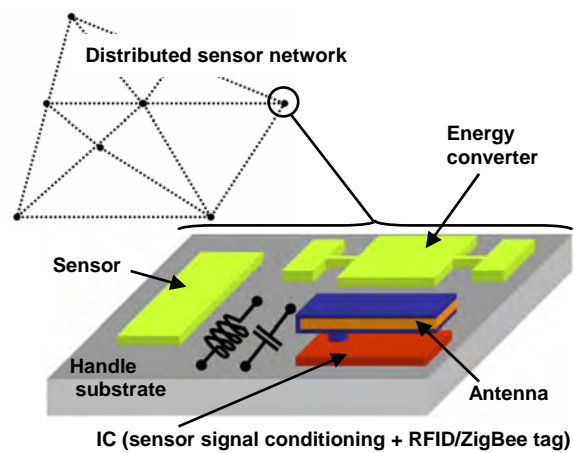

**Fig. 1 -** Project Overview: the vibration-to-electricity converter is the power source for nodes in a distributed network. Each node includes a sensor, a chip-size antenna [4] and an IC.

©EDA Publishing/DTIP 2007                                                                        ISBN: 978-2-35500-000-3

## 2. CONVERTER DESIGN

Vibrations with amplitude below 100 µm are often converted into electric energy thanks to a resonant comb-drive variable capacitor, with either In-Plane Gap-closing Combs (IPGC) [6] or In-Plane Overlap Combs (IPOC) [9]. Another type of variable capacitor, based upon simple parallel plate architecture, is Out-of-Plane Gap-Closing Plates (OPGP) [10]. Instead, recently some works have been reported to use two overlapping sets of planar electrodes with in-plane motion [11, 12, 1]. Hence this topology is termed as In-Plane Overlap Plates (IPOP). 3-D views of the presented variable capacitors design, based on this topology, are shown in Fig. 2. The devices are based on a silicon-glass technology [13]. IPOP resembles the concept of traditional parallel plate capacitor, with one set of electrodes patterned on the underside of silicon proof mass and other on top of glass wafer. By virtue of mechanical vibrations present in the environment, proof mass oscillates laterally. Hence the capacitance varies due to a change in the electrode's overlap area.

The IPOP topology has the following advantages. Large value of maximum capacitance is achievable, due to small air gap between the sets of top and bottom electrodes, leading to high power generation. The complexity of the fabrication process is reduced, as the proposed designs do not have any closely spaced comb drive architecture. So tolerance of misalignment is relatively higher as compared to other topologies which have comb drive variable capacitor. The IPOP also exhibits less fluid damping: as there is no compression of air, squeeze film damping is almost negligible.

The basic difference between the two proposed designs (cf. Fig. 2) is their proof mass. In the comb configuration (CC) 19 % of the total mass of silicon is removed from the zone present in between the electrodes to reduce the parasitic capacitance. It results in reduction of overall mass of the device, hence increasing its natural frequency as compared to the proof mass configuration (PC) of the same dimensions. Both converters have been designed to resonate at 300 Hz. Their dimensions are 11 x 6.5 x 0.9 mm$^3$, without including the connection pads. Vibration amplitude is limited to +/- 50 µm by the help of mechanical stoppers. This corresponds to the maximum capacitance variation. A 380 µm-thick silicon wafer is being used. Use of the thick wafer helps us in two ways. Firstly, net mass of the proof mass is increased, which helps us to reduce resonance frequency of the device, hence a low frequency vibration sources can be targeted. Secondly, out-of-plane vibrations are limited due to the high aspect-ratio springs, having 30 µm width and 380 µm height.

Fig. 3 shows FEM simulations of capacitive variation of the designs. Theoretically minimum capacitance $C_{min}$ should be equal to zero, but in real a parasitic capacitance is always present which is evident from the curves. The capacitance is composed of three components: the substrate parasitic capacitance $C_{sub}$, fringe field capacitance $C_{ff}$ between the two sets of electrodes and linear capacitance $C_{lin}$. To reduce $C_{min}$, $C_{sub}$ and $C_{ff}$ has to be minimized. $C_{min}$'s value in PC is quite higher as compared to CC. Reducing 19% of total mass from PC, results in 25% reduction of $C_{min}$ in CC. The non-linearities in the capacitance variations with displacements are due to the electrostatic fringe fields between top and bottom electrodes, as shown by the simulation of the capacitance's variation of face-to-face aluminum electrodes. $C_{ff}$ can be reduced by designing different electrode's width. However it is never desirable to have $C_{min}$ close to zero, because voltage will approach infinity and unwanted electrode stiction could occur. The last curve in Fig. 3 represents variation of $C_{lin}$. It is calculated by the following analytical formula:

$$C_{lin}(x) = N \frac{2\varepsilon_o \varepsilon_r L_F}{t_{NIT} + \varepsilon_r g_{AIR}} (W_F - x), \qquad (1)$$

where $N$, $\varepsilon_r$, $\varepsilon_o$, $L_F$, $W_F$, $t_{NIT}$, $g_{AIR}$, and $x$ are number of fingers, relative permittivity of dielectric, permittivity of free space, finger length, finger width, dielectric thickness, air gap present between two electrodes and lateral displacement of proof mass respectively.

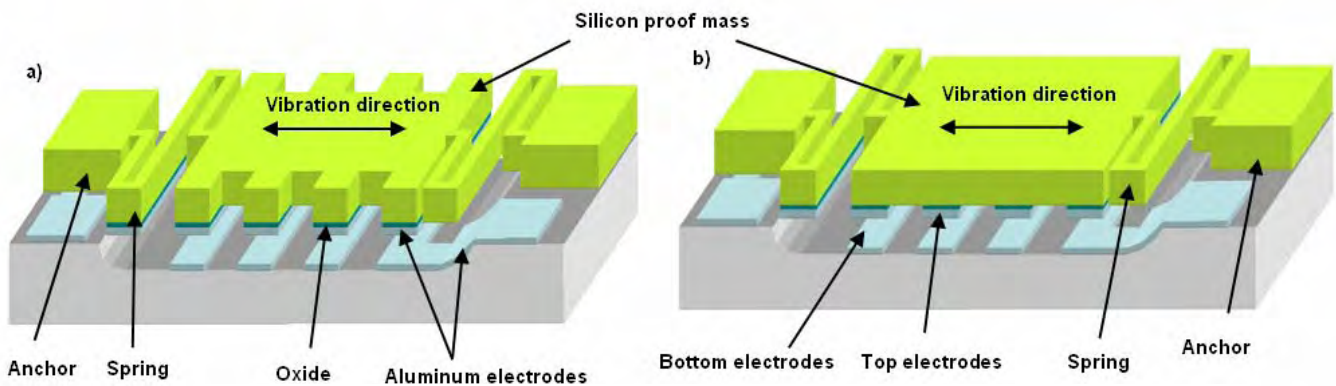

**Fig. 2** - Proposed designs: a) Proof mass configuration (PC), b) Comb configuration (CC)

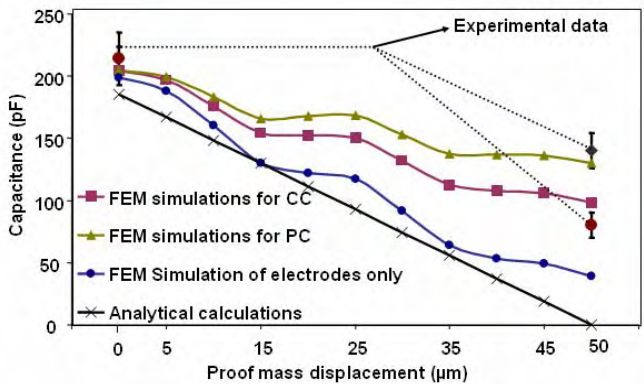

**Fig. 3** - FEM simulation of capacitance variation for the two presented designs

### 3. FABRICATION

The fabrication of the IPOP vibration-to-electricity converter involves parallel processing of the electrodes deposition on a silicon and on a glass wafer, after which both of them are anodically bonded. Ordinary cheap 380 µm-thick 4" silicon and 500 µm-thick glass wafers are used. The fabrication steps are shown in Fig. 4. Deposition of a thermal oxide on the silicon wafer insulates it from the aluminum electrodes. A nitride layer is deposited on lower electrodes, which are deposited on a glass wafer previously etched by HF, to serve as a passivation layer against unwanted stiction.

The most important issue is the through-wafer DRIE for silicon. At ESIEE, the silicon etching is performed using *Alcatel$^{tm}$* 601E plasma etcher. To achieve the vertical anisotropic deep plasma etching, the Bosch process is used. For our designs through-wafer silicon etching is required. But it is critical as our design has different beam widths as small as 30 µm and wide range of openings from 50 µm (gap between mechanical stopper and proofmass) to mm² holes (used to access metal contacts on a lower wafer). As thick silicon wafer is used, DRIE becomes particularly delicate because of the ARDE (Aspect Ratio Dependent Etching) and notching effects. The high aspect-ratio of the structures is obtained using a specific mixed RF/LF pulsed process allowing smooth sidewalls, perfect anisotropy and no negative effect of over-etching time [14]. Fig. 5 shows the microscopic view of the silicon wafer's backside after DRIE. It can be seen that the mechanical springs are fully released. The dark area shows the silicon dioxide from where the bulk silicon has been removed. Top view of the final device is shown in Fig. 6.

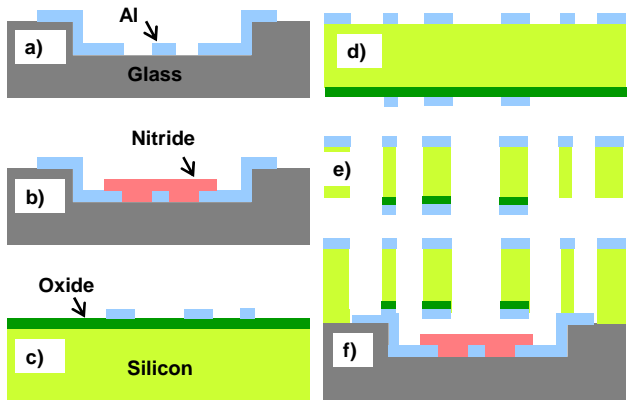

**Fig. 4 -** Fabrication process - a) glass etching + sputtering of Al, b) PECVD nitride passivation, c) thermal oxidation + sputtering of Al-electrodes on silicon back side, d) Al-sputtering of DRIE mask, e) through wafer DRIE, f) anodic bonding of substrate.

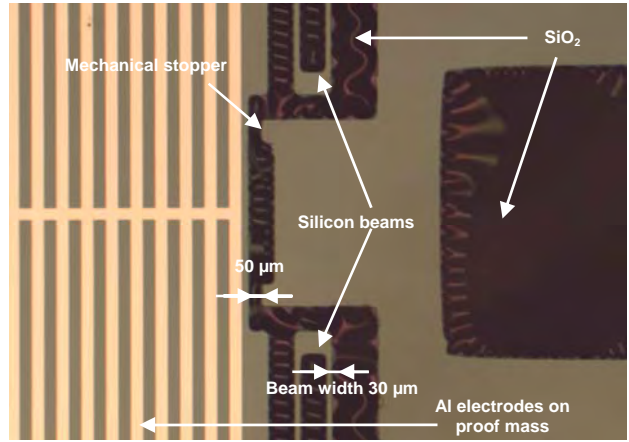

**Fig. 5** - Microscopic view of beams after DRIE of the silicon

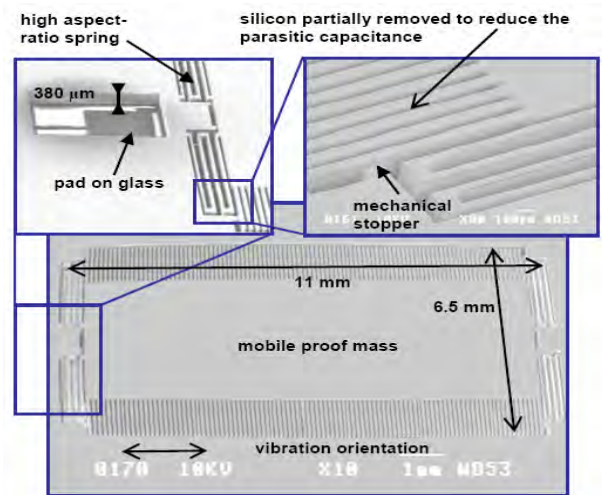

**Fig. 6** - SEM picture of the vibration-to-electricity converter after anodic bonding, with 2 close views

## 4. CHARACTERIZATION

The device's mechanical characterization has been done using a vibration table from *Physik Instrumente[tm]*. During the test, a sine wave is applied on the input of a dedicated current amplifier, generating a sinusoidal displacement of the table. While sweeping the frequency, the relative displacement of the structure's proof mass had been observed with an optical profiler from *Weeco[tm]*. For an unbounded silicon wafer, the resonance frequency of the PC and CC is found to be 307 Hz, which is in accordance with the simulations done in *Coventorware[tm]*. After the wafer bonding, the resonance frequency is found to be 290 Hz for the CC and 255 Hz for the PC. The decrease in resonance frequency is due to air being trapped in between the mobile and fixed parts after the bonding. The full wafer being excited with input amplitude of +/-5 µm, Fig. 7 shows, for the two designs, the maximum value of the relative displacement of proof mass versus frequency.

Experimental values of $C_{max}$ and $C_{min}$, when the potential of the silicon substrate is floating, are reported in Fig. 3. The maximum and minimum capacitances for the PC are 214 and 140 pF respectively. For CC, corresponding $C_{max}$ and $C_{min}$ are 214 and 80 pF. If substrate is grounded, most of the substrate parasitic capacitance is cancelled, so $C_{max}$ and $C_{min}$ in both cases are reduced by 33 pF.

The harvested energy, with an ideally lossless electronic, is given by the following equation [9]:

$$E = \frac{1}{2} V_{in}^2 (C_{max} - C_{min}) \left( \frac{C_{max}}{C_{min}} \right), \quad (2)$$

where $V_{in}$ is the starting voltage of the system. Power generated is calculated by multiplying the result of above equation with twice the resonance frequency, since in a single period the device capacitance reaches $C_{min}$ two times. For $V_{in}$ equals to 5 V, the harvested power would be 0.72 and 2.6 µW for the PC and CC respectively, when the potential of the silicon substrate is floating. Hence the power densities for PC and CC are 11.26 and 40.62 µW/cm³ respectively. When substrate is grounded, the corresponding harvested power for PC and CC becomes 0.82 µW and 3.74 µW. Thus resulting in a power density of 12.95 µW/cm³ and 59 µW/cm³, for PC and CC respectively.

## 5. POWER DENSITY IMPROVEMENT

In the CC 19% of total mass of silicon is removed as compared to the PC, resulting in decrease in the parasitic capacitance, hence giving a high generated power density. But this mass reduction undesirably increases the resonance frequency of the device. So spring stiffness should be reduced to make system apt for low frequency applications. As the resonance frequencies of the most useful vibration sources lie below 300 Hz. In order to achieve a system which can work at such a low frequency, we cannot afford to loose a huge mass. Hence some improvement is required to reduce the parasitic capacitance but without sacrificing lot of mass.

Electric field inside the silicon acts as a big parasitic capacitance. It is strong near the top electrodes when they are positioned at a minimum capacitance point, as there is a direct overlap of the silicon substrate and aluminum electrodes on glass wafer. So in order to reduce the parasitic capacitance, silicon between the electrodes having a high permittivity and being not highly resistive should be replaced by air. It can be done by through-wafer etching as a case of the CC but the problem is the loss of mass. To avoid this inconvenience, we propose to etch small depth of silicon present in between the top electrodes. This step is called as Backside DRIE. Fig. 8 shows a side view of modified design. Fig. 9 shows a variation in $C_{max}$ and $C_{min}$ as a function of backside DRIE depth, based on FEM simulations, using *Coventor$^{tm}$*. It can be concluded that through-wafer etching is not required. The value of $C_{min}$ decreases quite rapidly with an etched depth, and after 20 µm it almost becomes constant. At this point backside DRIE results in only 2.5% mass loss. The back side DRIE has no notable effect on the value of $C_{max}$. This minor mass change improves our capacitance variation, from 64 pF as in a case of the PC and 114 pF for the CC to 156 pF. Based on these FEM simulation results we can estimate an improvement in our power density. At a resonance frequency of 300 Hz with the loss less ideal electronic circuit, and a starting voltage of 5 V, the power density becomes 76.71 µW/cm$^3$. So an additional fabrication step i.e. backside DRIE of the silicon wafer has been added to our fabrication process. Before doing through-wafer top DRIE to release movable proof-mass, a depth of 20 µm silicon is etched, in a zone between the top electrodes. The remaining fabrication steps are exactly the same.

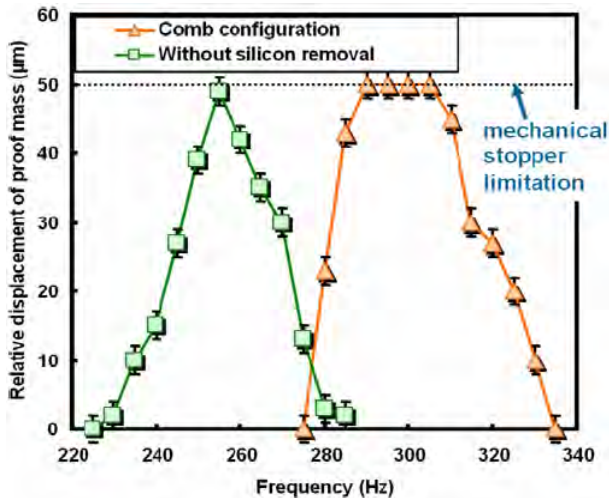

**Fig. 7** - Maximum relative displacement of proof mass versus frequency with sinusoidal input excitation

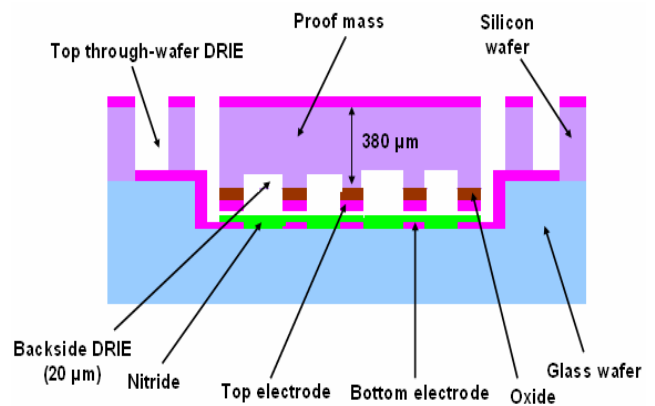

**Fig. 8** - Modified design with backside DRIE

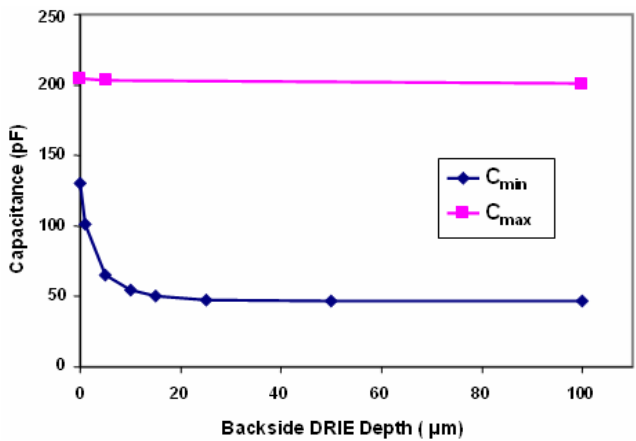

**Fig. 9** - Effect of backside DRIE on maximum and minimum capacitance (FEM simulation based result)

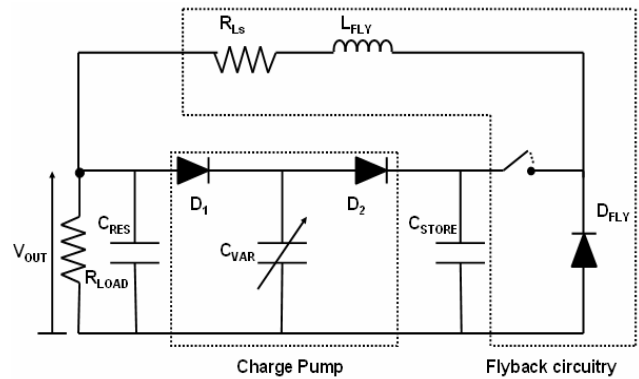

**Fig. 10** – Simulated electronic power processing circuit

## 6. POWER PROCESSING CIRCUITRY

For an implementation of the power processing electronics, we have chosen to optimize the electronic circuit described in [15]. It is composed of two main parts: charge pump and flyback circuitry, as shown in Fig. 10. The charge pump allows an accumulation of the electric charges. It is made up of two diodes $D_1$ and $D_2$, and the variable capacitor $C_{var}$. $D_1$ and $D_2$ function like switches, whereas $C_{var}$ is an energy harvesting source. When a sufficient quantity of charge is accumulated in the temporary storage $C_{store}$, it is then transferred periodically to the reservoir $C_{RES}$, using the flyback circuitry, whose configuration is same as of a buck chopper. This circuit includes a transistor represented by a switch, a free wheeling diode $D_{FLY}$ and an inductance $L_{FLY}$. One condition to be respected is $C_{RES} \gg C_{STORE}$ in order to maintain constant voltage across load.

There are three drawbacks associated with this architecture. Firstly it works well only at high resonance frequencies above about 1 kHz. Secondly for a too small value of $C_{min}$, there is no positive energy conversion. Thirdly there is a need of an additional circuit to control the switch in flyback circuitry.

For the switch we use a MOS transistor. The clock signal, used to make the transistor in a conduction state, has a significant influence on the circuit's power conversion efficiency. A set of simulations was performed to study the impact of the clock signal on the output voltage. Following values of different circuit components are kept fixed during simulations: $C_{RES}$ = 2 µF, $R_{Load}$ = 20 MΩ, $L_{FLY}$ = 4 mH, $V_{in}$ = 5 V, $C_{max}$ = 208 pF, $C_{min}$ = 47 pF and Freq = 300 Hz. Fig. 11 shows a relation of the clock's pulse width with the value of the output voltage and $C_{STORE}$. The optimum value of the pulse width is around 2 µsec with a value of $C_{STORE}$ around 2.2 nF. If the pulse width becomes too large, the differential voltage in the inductance becomes negative and with the transistor in a conduction state, the two diodes are on. As a result a short circuit is created and there is a no transfer of charge to $R_{LOAD}$. Fig. 12 shows the dependence of the output voltage on the value of a clock period. An optimum value of the clock period lies between 33 ms and 36 ms, which means ten energy conversion mechanical cycles for one feedback process. An optimum value of $C_{STORE}$ is 2.2 nF. Hence on the basis of these simulations we can conclude that the pulse width determines the fall of output voltage after one conduction cycle of the transistor whereas the clock period determines the peak of the output voltage. Simulations also show that for the pulse width higher then 1.5 µs, the output voltage is directly proportional to the size of

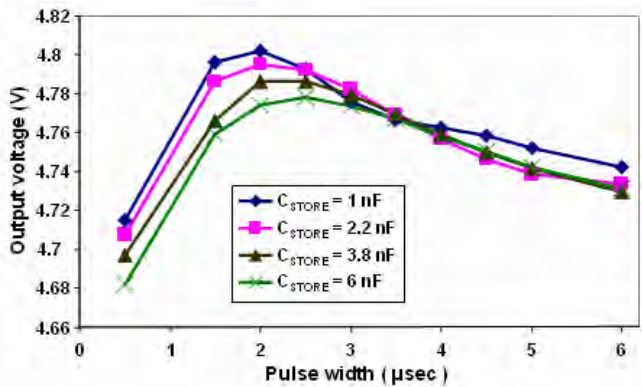

**Fig. 11**– Output voltage versus pulse width

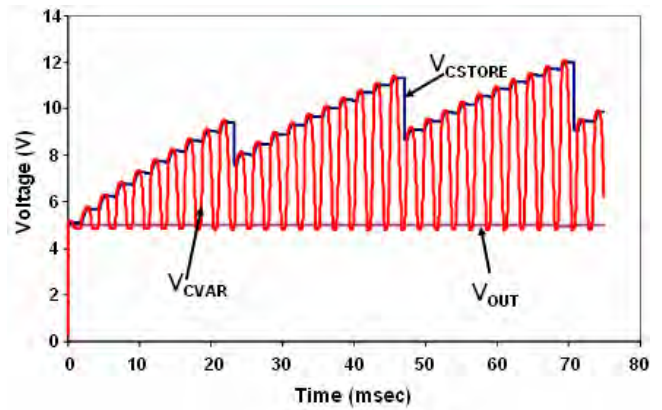

**Fig. 13**– $V_{CVAR}$, $V_{CSTORE}$ and $V_{OUT}$ versus time

### 7. CONCLUSION

We have presented a resonant electrostatic vibration-to-electricity converter using an In-Plane Overlap Plate topology. Batch CMOS-compatible process is performed on a typical silicon wafer that is eventually bonded with a glass wafer. High capacitance value, due to a vertical gap of 1.5 µm between the top and bottom electrodes, allows high energy conversion. Dimensions of the device without including the pads are 11 x 6.5 x 0.9 mm$^3$. Maximum and minimum capacitance values are 181 and 47 pF respectively, if the silicon substrate is grounded. The harvested power density would be 58 µW/cm$^3$ at 290 Hz, for a starting voltage of 5 V and a theoretically lossless electronic. The parasitic capacitance has a strong impact on the generated power. We have partially removed the silicon substrate from the proof mass by through-wafer DRIE, in the case of the CC, to reduce it. This leads to an increase in the power density but with a considerable mass inductance. Its value can be increased but this can cause a major integration problem.

Still we cannot have any optimized circuit with positive energy conversion for our device, as we are targeting very low frequencies. However it is observed that in order to have a circuitry that works well at low frequency the value of $C_{min}$ also needs an adjustment. For the circuit optimization at low frequencies all the circuit parameters have been kept the same as described above. The value of $C_{STORE}$ is chosen 2.2 nF, clock pulse width is kept 2 µsec and clock period is five times bigger then mechanical vibration period, based on above optimization results. Then frequency has been reduced and $C_{min}$ has been increased step by step. A successful simulation has been done for the vibration frequency of 410 Hz, and for the values of $C_{max}$ and $C_{min}$ of 368 pF and 94 pF respectively. The simulation results are shown in Fig. 13. $V_{CVAR}$ and $V_{CSTORE}$ are the voltages across $C_{VAR}$ and $C_{STORE}$ respectively. Fig. 14 is a zoom of $V_{OUT}$, which shows there is positive power conversion.

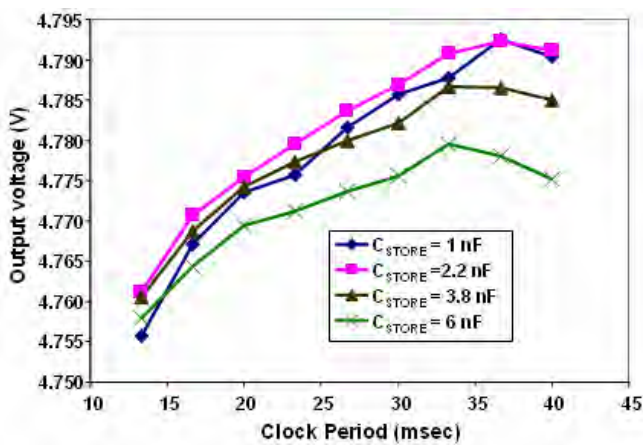

**Fig. 12**– Output voltage versus clock period

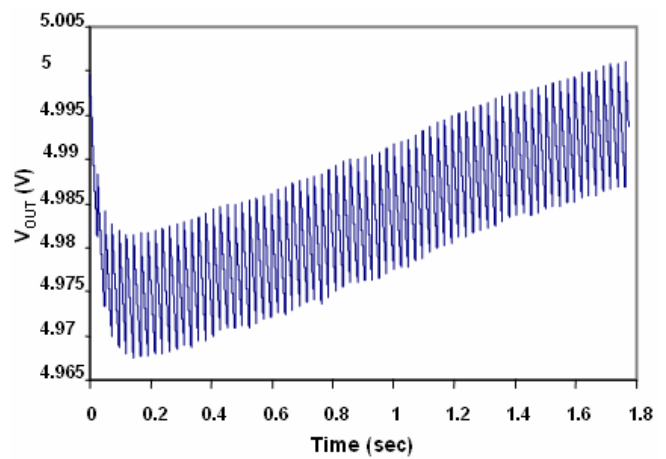

**Fig. 14**– Zoomed view of $V_{OUT}$

decrease. This is totally undesirable as the most common ambient vibrations are below 300 Hz. Hence we propose an enhancement in the fabrication process i.e. backside DRIE, which means to etch small silicon depth in a zone present between the top electrodes. It allows us to reduce parasitic capacitance without loosing a too much mass. By simulations we have shown that with a lossless electronics, starting voltage of 5 V, at a mechanical resonance frequency of 300 Hz and with Backside DRIE of 20 µm the power density becomes 76.71 µW/cm$^3$. It results only in 2.5 % mass loss as compared to 19 % mass loss in the CC. Finally a power processing electronic circuitry has been studied and corresponding problems for low frequency applications have been highlighted. Also an optimization of various parameters which effect the output voltage has been discussed.

## 8. ACKNOWLEDGEMENTS

Authors would like to thank the SMM team in ESIEE for their extensive cooperation during the fabrication process in the clean room. This work is partially funded by the French Research Agency (*ANR, Agence Nationale pour la Recherche*) and is supported by the French competitiveness cluster "*Ville et Mobilité Durable*".